\documentstyle[aps,preprint,floats]{revtex}

\newcommand{\beq}{\begin{equation}}
\newcommand{\eeq}{\end{equation}}
\newcommand{\beqa}{\begin{eqnarray}} 
\newcommand{\eeqa}{\end{eqnarray}}

\newcommand{\disp}{\displaystyle}
\def\sm{Standard Model}

\def\etal{{\it et al.}}
\newcommand{\pks}{\mbox{$a_{CP}(\psi K_S)$}}
\newcommand{\fks}{\mbox{$a_{CP}(\phi K_S)$}}
\newcommand{\stb}{\mbox{$\sin 2\beta$}}

\def\plb#1{Phys.\ Lett.\ {\bf B #1}}
\def\prd#1{Phys.\ Rev.\ {\bf D #1}}
\def\prl#1{Phys.\ Rev.\ Lett. {\bf #1}}

\def\zpc#1{Z.~Phys.\ {\bf C #1}}

\begin{document}

\draft

{\tighten

\preprint{
\vbox{
      \hbox{SLAC-PUB-7614}
      \hbox{DOE/ER/41014-27-N97}
      \hbox{hep-ph/9708305}
      \hbox{August 1997} }}

\renewcommand{\thefootnote}{\fnsymbol{footnote}}

\title{CP Asymmetry in $B_d \to \phi K_S$: \\
Standard Model Pollution
\footnotetext{Research at SLAC supported
by the Department of Energy under contract DE-AC03-76SF00515}}
\author{Yuval Grossman${}^a$, Gino Isidori${}^{a,b}$ and  
Mihir P. Worah${}^{a,c}$}
\address{ \vbox{\vskip 0.truecm}
${}^a$ Stanford Linear Accelerator Center, \\
        Stanford University, Stanford, CA 94309, USA \\
${}^b$INFN, Laboratori Nazionali di Frascati, 
                I-00044 Frascati, Italy \\
${}^c$ Department of Physics, \\
         University of Washington, Seattle, WA 98185, USA}

\maketitle
\thispagestyle{empty}
\setcounter{page}{0}
\begin{abstract}

The difference in the time 
dependent CP asymmetries between the modes $B \to \psi K_S$ and $B \to
\phi K_S$ is a clean signal for physics beyond the Standard
Model. This interpretation could fail if there is a large enhancement
of the matrix element of the $b \to u \bar u s$ operator 
between the $B_d$ initial state and the $\phi K_S$ final state. 
We argue against this possibility and propose some
experimental tests that could shed light on the situation. 

\end{abstract}

} 

\newpage

{\noindent \bf 1.}
It is well known that in the \sm\ the time--dependent CP--violating
asymmetry in  $B_d \to \psi K_S$ [\pks] measures \stb, where
$\beta=\arg (-{V_{cd}V_{cb}^*}/{V_{td}V_{tb}^*})$
and $V_{ij}$ denote the CKM matrix elements \cite{yossi,pdg}. 
Moreover, being dominated by the tree--level transition
$b\to c \bar c s$, the decay amplitude of $B_d \to \psi K_S$ 
is unlikely to receive significant corrections from 
new physics.\footnote{
There is, of course, a possible new contribution to 
the $B^0-\bar B^0$ mixing amplitude. This does not 
affect the generality of our arguments or the conclusions \cite{us}.}
Interestingly, within the 
\sm\ the CP  asymmetry in $B_d \to \phi K_S$ [\fks] 
also measures \stb\ if,  
as naively expected,
the decay amplitude is dominated by the
short--distance penguin transition $b\to s\bar s s$ \cite{lpgg}. 
Since $B_d \to \phi K_S$ is a loop mediated 
process within the \sm, 
it is not unlikely that new physics could
have a significant effect on it \cite{us}. 
The expected branching ratio and the high identification
efficiency for this decay suggests 
that \fks\ is experimentally accessible at the early stages of 
the asymmetric $B$ factories. 
Thus, the search for a difference between \pks\ and \fks\ is a 
promising way to look for physics beyond the \sm\ 
\cite{us,n-q,ciuchini,london,barbieri}. 

If, indeed, it turns out that \pks\ is not equal to \fks,
it would be extremely important to know how
precise the \sm\ prediction of them being equal is. In particular, one
has to rule out the possibility of unexpected long distance effects
altering the prediction that \fks\ measures \stb\ in the
\sm.

The weak phases of the
transition amplitudes are ruled by products of CKM matrix elements. 
In the $b \to s q \bar q$ case,
relevant to both $B_d \to \psi K_S$ and $B_d \to \phi K_S$,
we denote these by 
$\lambda^{(s)}_q = V_{qb}V_{qs}^*$.
For the purpose of CP violation studies, it is instructive to 
use CKM unitarity and express any 
decay amplitude as a sum of two terms \cite{gq}. In particular,
for $b \to s q \bar q$ we eliminate $\lambda^{(s)}_t$ and write
\beq \label{twoterms-s}
A_f =  \lambda^{(s)}_c A_f^{cs} + \lambda^{(s)}_u A_f^{us}.
\eeq
The unitarity and the experimental hierarchy of the CKM matrix
imply \cite{wolf} $\lambda^{(s)}_t \simeq \lambda^{(s)}_c
\simeq A\lambda^2
+ {\cal O}(\lambda^4)$ and $\lambda^{(s)}_u = A\lambda^4 e^{i\gamma}$,
where $A \approx 0.8$, $\lambda=\sin\theta_c=0.22$ and
$\gamma$ is a phase of order one.
Thus the first and dominant term is real 
(we work in the standard parametrization).
The correction due to the second term, that is complex and  
doubly Cabibbo suppressed, is 
negligibly small unless $A_f^{us} \gg A_f^{cs}$.

The $A_f^{qs}$ amplitudes cannot be calculated exactly since they  
depend on hadronic matrix elements. 
However, in some cases we can reliably estimate their relative sizes.
For $B \to \psi K_S$ the
dominant term includes a tree level diagram while the CKM--suppressed term 
contains only one--loop (penguin) and higher order diagrams.
This leads to  
$A_{\psi K_S}^{cs} \gg A_{\psi K_S}^{us}$, and thus 
insures that \pks\ measures \stb\ in the \sm.
Since both terms for $B \to \phi K_S$ begin at one-loop order one
naively expects  $A_{\phi K_S}^{cs} \sim A_{\phi K_S}^{us}$. In this case 
\fks\ also measures \stb\ in the \sm\ up to corrections of 
${\cal O}(\lambda^2)$.  However, any unexpected enhancement 
of $A_{\phi K_S}^{us}$ would violate this result.
In particular, 
an enhancement of ${\cal O}(\lambda^{-2}) \sim 25$ (analogous
to the $\Delta I = 1/2$ rule in $K$ decays) leads to ${\cal O}(1)$
violations, and subsequently 
to $\pks \ne\fks$ even in the \sm. 

In this note we argue against this possibility,
presenting different arguments that suggest the pollution of 
$A_{\phi K_S}^{us}$ in $B_d \to \phi K_S$ is very small. 
Moreover, we will propose some experimental tests 
that in the near future could provide quantitative bounds on this
pollution. 

\bigskip
{\noindent \bf 2.}
The natural tool to describe the $B$ decays of interest is
by means of an effective $b\to s \bar q q$ Hamiltonian.
This can be generally written as 
\beq
{\cal H}_{eff}^{(s)} = \frac{G_F}{\sqrt{2}}\left\{ 
\lambda^{(s)}_t \sum_{k=3..10} C_k(\mu) Q^{s}_k +
\lambda^{(s)}_c \sum_{k=1,2} C_k(\mu) Q^{cs}_k +
\lambda^{(s)}_u \sum_{k=1,2} C_k(\mu) Q^{us}_k 
\right\}\,, 
\label{Heff}
\eeq
where $Q_k^i$ denote the local four fermion operators  and $C_k(\mu)$ 
the corresponding Wilson coefficients, to be 
evaluated at a renormalization scale $\mu\sim {\cal O}(m_b)$.
For our discussion it is useful 
to emphasize the flavor structure of the operators:
$Q^{qs}_{1,2} \sim \bar b s \bar q q$  and $Q^{s}_{3..8} \sim \bar b s
\disp{\sum_{q=u,d,s,c}} \bar q q$, as well as the order of  
magnitude of their Wilson coefficients: 
$C_{1,2}\sim {\cal O}(1)$ and $C_{3..8}\sim {\cal O}(10^{-2})$.
The estimates of the $C_k(\mu)$ beyond the leading logarithmic 
approximation and the definitions of the $Q^i_k$, 
can be found in \cite{buras}. 
To an accuracy 
of $O(\lambda^2)$  in the weak phases, ${\cal H}_{eff}^{(s)}$
can be rewritten as 
\beq \label{approxh}
{\cal H}_{eff}^{(s)} = \frac{G_F}{\sqrt{2}}\left\{ 
\lambda^{(s)}_c \left[ \sum_{k=1,2} C_k(\mu) Q^{cs}_k -
\sum_{k=3..10} C_k(\mu) Q^{s}_k \right] +
\lambda^{(s)}_u \sum_{k=1,2} C_k(\mu) Q^{us}_k \right\}~.
\eeq
It is clear that, when sandwiched between the 
$B_d$ initial state and the $\phi K_S$ final state,
the first term corresponds to 
$A_{\phi K_S}^{cs}$ and the second to $A_{\phi K_S}^{us}$ [{\it cf}
Eq. (\ref{twoterms-s})]. The pollution is then generated by 
$Q^{us}_{1,2}$, corresponding to the $b \to s \bar u u$ transition.

Since the matrix elements of the $Q_k^i$ have to be evaluated 
at $\mu \sim {\cal O}(m_b)$, a realistic estimate of their relative 
sizes can be obtained within perturbative QCD.
We recall that  the $|\phi\rangle$ is an almost pure $|\bar s s\rangle$ state. 
The $\omega-\phi$ mixing angle is estimated 
to be below $5\%$ \cite{phimix,pdg}. 
We neglect this small mixing in the following.
Then, the matrix elements of $Q_{1,2}^{us}$ and $Q_{1,2}^{cs}$
evaluated at the leading order (LO) in the factorization
approximation are identically zero. At LO only $Q_{3..8}$,
i.e. the short--distance $b \to s \bar s s$ penguins, have a non 
vanishing matrix element in $B_d \to \phi K_S$. As a consequence, 
the weak phase of the $B_d \to \phi K_S$ decay amplitude is essentially zero.
Nonetheless, given the large Wilson coefficients of $Q^{qs}_{1,2}$, a 
more accurate estimate of their contribution is required.

At next to leading order (NLO), working in a modified factorization 
approximation, one obtains additional contributions from penguin--like matrix
elements of the operators $Q_{2}^{us}$ and $Q_{2}^{cs}$
\cite{fleisch1}. These have been reevaluated recently, and
shown to be important in explaining the CLEO data on charmless
two--body B decays \cite{ciuchini2,fleischer,lenz}. However, 
even in this case the $b \to s \bar u u$ 
pollution in $B_d \to \phi K_S$ is very small.
The reason is that, in the limit where we can neglect both the 
charm and the up quark masses with respect to $m_b$, the 
matrix elements of $Q^{us}_{1,2}$ and $Q^{cs}_{1,2}$ 
are identical from the point of view of perturbative QCD (up to
corrections of ${\cal O}(m_c/m_b)\sim 0.3$).
However, the overall contribution of the charm operators $Q^{cs}_{1,2}$ 
is enhanced by a factor $\lambda^{-2}$
with respect to the one of $Q^{us}_{1,2}$.
Thus, either if the $B_d \to \phi K_S$ transition is dominated by 
$Q^{s}_{3-10}$ (short--distance penguins) or if it is 
dominated by $Q^{cs}_{1,2}$ (long--distance charming penguins),
the weak phase is vanishingly small.

Of course one could
not exclude a priori a scenario where the contributions of 
$Q^{s}_{3..8}$ and $Q^{cs}_{1,2}$ cancel each other to 
an accuracy of $O(\lambda^2)$. However, this
extremely unlikely possibility would result in an unobservably small  
$BR(B_d \to \phi K_S)$, rendering this entire discussion moot.

As discussed above, any
enhancement of $\langle \phi K_S |Q^{us}_{1,2}  | B_d \rangle$, 
that could spoil the prediction that \fks\ measures \stb\ in the \sm\
should occur at low energies in order not to be compensated
by a corresponding enhancement of 
$\langle \phi K_S | Q^{cs}_{1,2} | B_d \rangle$.
This possibility is not only disfavored  by the OZI rule 
\cite{ozi},\footnote{This non--perturbative prescription has never 
been fully understood in the framework of perturbative QCD, but
can be justified in the framework of the $1/N_c$ expansion, 
and is  known to work well in most cases and
particularly in the vector meson sector \protect\cite{isgur}.}
but is also 
suppressed by the smallness of the energy range where 
the enhancement should occur with respect to the scale of the process. 
We are not aware of any dynamical
mechanism that could favor this scenario. 
Inelastic rescattering effects in $B$ decays due to
Pomeron exchange have been argued not to be negligible
and to violate the factorization limit \cite{donoghue}. However,
even within this context violations of the OZI rule are 
likely to be suppressed \cite{ozi-viol}.

\bigskip
{\noindent \bf 3.}
There are experimental tests of our arguments that can be 
achieved in the sector of $b\to d$ transitions. 
These are described 
by an effective Hamiltonian ${\cal H}_{eff}^{(d)}$ completely similar
to the one in Eq. (\ref{Heff}) except for the substitution 
$s\to d$ in the flavor indices of both CKM factors and four--fermion 
operators.
$SU(3)$ flavor symmetry can be used to obtain relation among several
matrix elements. In particular
\beq
\sqrt{2}\;\langle \phi K_S  |Q^{us}_{1,2}  | B_d \rangle =
\langle \phi\pi^+ |Q^{ud}_{1,2}  | B^+ \rangle +
\langle K^* K^+|Q^{ud}_{1,2}  | B^+ \rangle\,.
\eeq
($SU(3)$ breaking effects, which are typically at
the 30\% level, are neglected here.)
The coefficients of these matrix elements are, 
however, proportional to  different CKM factors.
This is illustrated in Table I, where we show the relevant $B$ decay
modes along with the Cabibbo factors corresponding to the leading
and sub--leading contributions to the decay amplitudes.
If our arguments hold, one expects  $BR(B_d \to \phi K_S) 
\sim {\cal O}(\lambda^4)$ and $BR(B^+ \to K^*K^+)$,
$BR(B^+ \to \phi \pi^+) \sim {\cal O}(\lambda^6)$. Notice, however,  that
the overall contribution of $Q^{ud}_{1,2}$ in 
$B^+ \to K^* K^+$  and $B^+ \to \phi \pi^+$ is enhanced with
respect to the one of $Q^{us}_{1,2}$ in $B_d \to \phi K_S$
by the corresponding CKM factors: $\lambda_u^{(d)}/\lambda_u^{(s)} 
= {\cal O}(\lambda^{-1})$. 
Thus, if $\langle \phi K_S  |Q^{us}_{1,2}  | B_d \rangle$
is enhanced by ${\cal O}(\lambda^{-2})$ in order
to interfere with the dominant 
${\cal O} (\lambda^2)$ contributions, then  $BR(B^+ \to \phi \pi^+)$ 
and/or $BR(B^+ \to K^*K^+)$ would be dominated by the similarly enhanced 
matrix elements of $Q^{ud}_{1,2}$. 
This would result in an enhancement of 
the naively Cabibbo suppressed modes, 
i.e. we should observe $BR(B^+ \to \phi \pi^+) \sim {\cal O}(\lambda^2)$ 
and/or $BR(B^+ \to K^*K^+) \sim {\cal O}(\lambda^2)$
[while $BR(B_d \to \phi K_S)$ is still $\sim {\cal O}(\lambda^4)$].
Similar arguments hold for the corresponding $B_d$ decay modes,
however in that case the $SU(3)$ relation is not quite as precise.

\begin{table}
\[ \begin{array}{||c||c|c|c||} \hline\hline
\qquad\quad\mbox{Decay\ mode}\qquad\quad & \multicolumn{3}{c||}{
     \mbox{Operators \ and \ CKM\ factors}   }  \\
&\quad \mbox{penguins}\quad &\quad c\mbox{--trees}\quad &
 \quad u\mbox{--trees}\quad  \\ \hline
B_d \to \phi K_S   & Q^s_{3..8}    & Q^{cs}_{1,2}   & Q^{us}_{1,2}  \\ 
                 & \lambda_t^{(s)} \sim \lambda^2 
                 & \lambda_c^{(s)} \sim \lambda^2 
	         & \lambda_u^{(s)} \sim \lambda^4 \\  \hline
B^+ \to \phi \pi^+ \ {\rm and} \ B^+ \to K^*K^+
                 & Q^d_{3..8}    & Q^{cd}_{1,2}   & Q^{ud}_{1,2}  \\
                 & \lambda_t^{(d)} \sim \lambda^3 
                 & \lambda_c^{(d)} \sim \lambda^3 
	         & \lambda_u^{(d)} \sim \lambda^3 \\ \hline\hline
\end{array} \]
\label{su3}
\vspace{0.1in}
\caption
{$SU(3)$ related $B$ decay modes that allow us to quantify the \sm\
pollution in \fks.}
\end{table}

To get a quantitative bound we define the ratios
\beq
R_1 = \frac{BR(B^+ \to \phi \pi^+ )}{BR(B_d \to \phi K_S)}, \qquad
R_2 = \frac{BR(B^+ \to K^*K^+ )}{BR(B_d \to \phi K_S)}\,,
\eeq
such that in the \sm\ the following inequality holds
\beq 
\left| \pks-\fks \right|
< \frac{ \lambda}{\sqrt{2}} \left( \sqrt{R_1} + \sqrt{R_2} \right) 
[1+R_{SU(3)}] + {\cal O}(\lambda^2)\,,
\label{limit}
\eeq
where $R_{SU(3)}$ represents the $SU(3)$ breaking effects.
While measuring \fks\ it should be possible to 
set limits at least of order one on $R_1$ and $R_2$ and
thus to control by means of Eq. (\ref{limit}) the accuracy to 
which \fks\ measures $\sin 2\beta$ in the \sm. The limits
$\sqrt{R_1}, \sqrt{R_2} \lesssim 0.25$ would reduce the theoretical
uncertainty to the $10\%$ level.

It may be possible to confirm that $BR(B^+ \to \phi \pi^+ )$ and
$BR(B^+ \to K^*K^+ )$ are not drastically enhanced based just on the
current CLEO data. The CLEO colloboration already has reported the
bounds  
$BR(B^+ \to \phi K^+ ) < 1.2\times 10^{-5}$ and  
$BR(B^+ \to K^*\pi^+ ) < 4.1\times 10^{-5}$ \cite{cleo}. 
Given the similarity of energetic $K$'s and $\pi$'s in the CLEO
environment,  
it is plausible that similar bounds can also be derived for the modes
$B^+ \to \phi \pi^+$ and $B^+ \to K^*K^+$ respectively. A specialized
study of these modes is currently under way. Bounds on these branching
ratios of ${\cal O}(10^{-5})$ would clearly imply that the rates are
not ${\cal O}(\lambda^2)$ as they would be if the matrix elements of
$Q^{ud}_{1,2}$ were enhanced by ${\cal O}(\lambda^{-2})$.

The above experimental test can only confirm that \fks\ measures
\stb\ in the \sm. If it turns out that 
$R_1$ or $R_2$ is large, this may be either
due to the failure of our conjectures or due to new physics. 
If, however, $R_1$ and $R_2$ are small, and  
$\pks - \fks$ violates the \sm\ prediction of Eq. (\ref{limit}), this
would be an unambiguous sign of new physics.

Another possible check of our conjecture could be achieved 
through the measurement of the CP asymmetry in 
$B_d \to \eta' K_S$. Recently CLEO has measured a large branching 
ratio for the related decay $B^+ \to \eta' K^+$, suggesting these
processes are penguin dominated and thus that $a_{CP}(\eta' K_S)$ 
also should measure $\sin2\beta$ in the \sm\ \cite{london}. 
Nonetheless, the $|\eta'\rangle$ has a non negligible 
$|\bar u u\rangle$ component that could enhance the 
$b \to u\bar u s$ pollution and the $\eta'$ mass is 
one of the few exception where the OZI rule 
is known to be badly broken. Thus, without fine tuning, a sufficient 
condition to support our claim on \fks\ could be 
obtained by an experimental evidence of
$a_{CP}(\eta' K_S)=a_{CP}(\phi K_S)$. This would imply that the 
$b \to u\bar u s$ pollution is negligible in both cases.

\bigskip
{\noindent \bf 4.}
To summarize, we have argued that the deviation from the prediction that 
\fks\ measures $\sin 2\beta$ in the \sm\ is 
of ${\cal O}(\lambda^2) \sim 5\%$.
Moreover, we have shown how the accuracy of this prediction
can be tested experimentally.
While we concentrated on the time-dependent CP asymmetry it is clear that 
our arguments hold also for direct CP violation in charged and neutral
$B \to \phi K$ decays. Namely, that 
in the \sm\ the direct CP asymmetry is ${\cal O}(\lambda^2)$. 
Experimentally, we can hope to get an accuracy for both the time dependent
and the direct CP violation of about $10\%$.
Therefore, any measurable direct CP violation in $B \to \phi K$
or an indication that $\pks \ne \fks$,
combined with experimental evidence that the \sm\ pollution is of 
${\cal O}(\lambda^2)$
will signal physics beyond the \sm.

\bigskip
{\noindent \bf \it Acknowledgments.}
We thank Ann Nelson for instigating this investigation.
We have benefitted from useful conversations with Roy Briere, Gerhard
Buchalla, Bob Cahn, Alex Kagan, 
David Kaplan, Marek Karliner, Sacha Kop and Yossi Nir.

{\tighten

} 


\begin{references}

\bibitem{yossi}
For a review see e.g.
Y. Nir, Lectures presented in the 20th SLAC Summer Institute,
SLAC-PUB-5874 (1992); 
Y. Nir and H.R. Quinn, Ann. Rev. Nucl. Part. Sci. {\bf 42}, 211 (1992).

\bibitem{pdg}
R.M. Barnett {\it et al.}, the Particle Data Group, \prd{54}, 1 (1996).


\bibitem{us} 
Y. Grossman and M. Worah, \plb{395}, 241 (1997), and in 2nd
International Conference on B Physics and CP violation,
hep-ph/9707280. 

\bibitem{lpgg}
D. London and R.D. Peccei, \plb{223}, 257 (1989).

\bibitem{n-q}
Y. Nir and H. Quinn, \prd{42}, 1473 (1990).

\bibitem{ciuchini}
M. Ciuchini \etal, CERN-TH-97-047, hep-ph/9704274.

\bibitem{london}
D. London and A. Soni, UDEM-GPP-TH-97-40, hep-ph/9704277.

\bibitem{barbieri}
R. Barbieri and A. Strumia, IFUP-TH-16-97, hep-ph/9704402.


\bibitem{gq}
Y. Grossman and H.R. Quinn, hep-ph/9705356.

\bibitem{wolf}
L. Wolfenstein, \prl{51} 1945 (1983).

\bibitem{buras}
A. Buras and R. Fleischer TUM-HEP-275-97, hep-ph/9704376.

\bibitem{phimix}
R.S. Chivukula and M. Flynn, \plb{186}, 127 (1986);
P. Jain \etal, \prd{37}, 3252 (1988).

\bibitem{fleisch1}
R. Fleischer, \zpc{58}, 483 (1993).

\bibitem{ciuchini2}
M. Ciuchini \etal, CERN-TH-97-030, hep-ph/9703353; CERN-TH/97-188,\\
hep-ph/9708222. 

\bibitem{fleischer}
R. Fleischer and T. Mannel, TTP97-17, hep-ph/9704423.

\bibitem{lenz}
A. Lenz, U. Nierste and G. Ostermeyer; MPI-PH-97-015, hep-ph/9706501.

\bibitem{ozi} 
S. Okubo, Phys. Lett. {\bf 5}, 165 (1963); G. Zweig, CERN Reports 
TH-401 and TH-412 (1964); J. Iiuzuka, Prog. Theor. Part. Sci. 
{\bf 37}, 21 (1996).

\bibitem{isgur}  P. Geiger and  N. Isgur, \prl{67}, 1066 (1991); 
\prd{47}, 5059 (1993).


\bibitem{donoghue}
J. Donoghue \etal, \prl{77}, 2178 (1996).

\bibitem{ozi-viol} See e.g. G. Veneziano in ``Color Symmetry and
Quark Confinement'', ed. T. Thanh Van, Proceedings of the 12th 
Rencontr\'e de Moriond (1977). 

\bibitem{cleo}
The CLEO collaboration, D. Asner \etal, \prd{53}, 1039 (1996).


\end{references}
\end{document}